# Specifying Data Bases Management Systems by Using RM-ODP Engineering Language

Jalal laassiri[1], Said elhajji[2], Mohamed bouhdadi[3], Ghizlane Orhanou[4] and Youssef balouki[5]

[1] University Mohammed V-Agdal, Faculty of Sciences, Department of Mathematic and Informatics, Laboratory of Mathematic and Informatics and Applications, Rabat, BP 1014, Morocco

[2] University Mohammed V-Agdal, Faculty of Sciences Rabat

[3] University Mohammed V-Agdal, Faculty of Sciences Rabat

[4] University Mohammed V-Agdal, Faculty of Sciences Rabat

[5] University Mohammed V-Agdal, Faculty of Sciences Rabat

**Abstract**
Distributed systems can be very large and complex. The various considerations that influence their design can result in a substantial specification, which requires a structured framework that has to be managed successfully. The purpose of the RM-ODP is to define such a framework. The Reference Model for Open Distributed Processing (RM-ODP) provides a framework within which support of distribution, inter-working and portability can be integrated. It defines: an object model, architectural concepts and architecture for the development of ODP systems in terms of five viewpoints. Which include an information viewpoint.
Since the usage of Data bases management systems (DBMS) in complex networks is increasing considerably, we are interested, in our work, in giving DBMS specifications through the use of the three schemas (static, dynamic, invariant).
The present paper is organized as follows. After a literature review, we will describe then the subset of concepts considered in this work named the database management system (DBMS) object model. In the third section, we will be interested in the engineering language and DMBS structure by describing essentially DBMS objects. Finally, we will present DBMS engineering specifications and makes the connection between models and their instances. This introduces the basic form of the semantic approach we have described here.

## 1. State of the art
### 1.1. RM-ODP overview

The rapid growth of distributed processing has led to a need of coordinating framework for the standardization of Open Distributed Processing (ODP).
The open distributed processing (ODP) computational viewpoint describes the functionality of a system and its environment, in terms of a configuration of interacting objects at system interfaces, independently of their distribution. In addition, Quality of service (QoS) contracts and service level agreements are an integral part of any computational specification, which is specified in ODP in terms of environment contracts.
The Reference Model for ODP (RM-ODP) is a framework for the construction of open distributed systems [1]-[4]. It creates an architecture supporting distribution, networking and portability.
The foundations part [2] contains the definition of concepts and analytical framework for normalized description of (arbitrary) distributed processing systems. These concepts are gathered in several categories including basic modeling concepts, specification concepts, organizational concepts, and structuring concepts.
The architecture part [3] contains specifications of the required characteristics that qualify distributed processing to be open. It defines a framework containing:
Five viewpoints called: enterprise, information, computation, engineering and technology; which provide a basis for the ODP systems specification.
A language for each viewpoint, defining concepts and rules to specify ODP systems from the corresponding viewpoint.
Specifications of functions required to support ODP systems.
Transparency prescriptions, showing how to use the ODP functions to achieve distribution transparency.
In other words, the first three viewpoints points do not take into account neither distribution nor heterogeneity inherent problems. This principle corresponds closely to the concepts of PIM (Platform Independent Model) and PSM (Platform Specific Model) models in MDA (Model

Driven Architecture) architecture [5]. However, RM-ODP is a meta-norm [6] and can not be directly applied. Indeed, for instance, the viewpoint languages are abstract in sense that they define what concepts should be supported, not how these concepts should be represented. It is important that RM-ODP does not use the term language in its largest sense: a set of terms and rules for the construction of statements from terms; it does not propose any notation for supporting viewpoint languages.

In fact, RM-ODP provides only a framework for the definition of new ODP standards. These standards include the following standards:

- Standards for ODP functions [7],[8];
- Standards for modeling and specifying ODP systems;
- Standards for methodology, programming, implementing, and testing ODP systems.

Elsewhere, the languages Z [9], SDL [10], LOTOS [11] and, Esterelle [12] are used in RM-ODP architectural semantics part [4] for the specification of ODP concepts. Unfortunately, up to now, no formal method is suitable to specify and verify every aspect of an ODP system. The inherent characteristics of ODP systems imply the need to integrate different specification languages and to handle non-behavioral properties of ODP systems that is the QoS concepts.

### 1.2. UML Meta-Model Adoption

There had been an amount of research to apply UML [13] as a syntactic notation for the ODP viewpoint language [14]-[16]. The taken approach is to give a meta-model description for the language; it is a definition of this language by itself. This is presented in terms of three views:

- The abstract syntax. It is expressed using a subset of UML static modeling notations that are class diagrams.
- The well-formedness rules. These rules are expressed in OCL [17], a precise language based on first order-logic. OCL is used for expressing constraints on objects structure which cannot be expressed by class diagrams only.
- And the modeling elements semantics. We used the meta-modeling approach [18] to define syntax of a sub-language for ODP QoS-aware enterprise viewpoint specifications.

Furthermore, a part of UML meta-model itself has a precise semantic [19] defined using denotational meta-modeling approach. The denotational approach [20] is realized by defining the instance form of every language element and a set of rules determining which instances are denoted or not by a particular language element. There are three main steps through a denotational meta-modeling approach to the semantics:

1. Define the meta-model for the model's language: object template, interface template, action template, type, and role.
2. Define the meta-model for the instances' language: objects, binders, and interfaces.
3. Define the mapping or the meaning function between these two languages.

There are good reasons for adopting the UML meta-modeling approach in context of ODP systems. The UML meta-models provide a precise core of any case tool. The tools include a consistency checker that makes sure that invariants defined on a model do not conflict. In fact, a consistency checker between meta-models insures that different system specifications are consistent and do not conflict.

Besides, for testing ODP systems [2]-[3], the current techniques [21],[22] are not widely accepted. A new approach for testing, named agile programming [23], [24] or test first approach [25] is being increasingly adopted. The opinion is integrating system model and testing model using UML meta-modeling approach [26]. This approach is based on the executable UML [27]. In this context, OCL is used to specify the properties that have to be tested. OCL also serves to attach constraints to UML meta-models in order to verify the coherence of meta-models and to translate the constraints into code to evaluate them on instance models.

The part of RM-ODP considered in this paper is a subset for describing and constraining the structure of ODP information viewpoint specifications. It consists of modeling and specifying concepts defined in the RM-ODP foundations part and concepts in the information language. The UML/OCL meta-model developed here elaborates the conceptual core of the ODP information viewpoint language. We do not consider concepts for describing dynamic behavior.

### 2. RM-ODP Presentation

As seen above, RM-ODP is a framework for the construction of open distributed systems. It defines a generic object model in foundations part, and an architecture which contains specifications of the required characteristics that qualify distributed processing as open. The architecture extends and specialized object concepts of foundations part. In addition, the RM-ODP architecture model consists of a set of five viewpoint models, the concepts and rules associated with the language of each model, the distribution transparency constructs, and the ODP functions. The entire RM-ODP model is based on the RM-ODP foundations of an object model, rules for specification, and rules for structuring RM-ODP (Model Reference - Open Distributed Processing) [ISO96a] [ISO96b] [ISO98] which is an international standard published by ISO/IEC. It provides a reference model for the specification of open distributed applications.

The concept of RM-ODP viewpoints framework, therefore, is to provide separate viewpoints into the specification of a given complex system. Each one of these viewpoints satisfies an audience which is interested in a particular set of aspects of the system. Associated with each viewpoint, there is a viewpoint language [14, 16] that optimizes the vocabulary and presentation for the concerned audience.

Furthermore, the RM-ODP model can describe a system according to five viewpoints and each viewpoint is interested in a particular aspect of the system. These viewpoints are the following:

- **Enterprise**: It introduces the necessary concepts to represent a system in the context of an enterprise on which it operates. It is interested in the objective and the policies of a system.
- **Information**: It is focused on the semantics of information and the treatment carried out on information. The information is extracted from individual entities and the viewpoint describes information sources, sinks, and flows. A system is then described by information objects, relationships and behavior. The description is expressed through the use of three diagrams named invariant, static and dynamic.
- **Computational**: It allows a functional decomposition of the system. The various functions are fulfilled by objects that interact thanks to their interfaces.
- **Engineering**: It is focused on the deployment and communication of a system. It defines communication concepts like channel, stub, skeleton and deployment concepts like cluster, capsule, etc.
- **Technology**: It describes the implementation of a system in term of configuration of technical objects representing the hardware and software components of the implementation.

After specifying above the different viewpoints, we want to note that, a viewpoint is a subdivision of the specification of a complete system, established to bring together those particular pieces of information relevant to some particular area of concern during the design of the system.

Moreover, the relations between the actors of a system are mediated by means of languages which depend on the position of these actors within the system, on their own activity (designer, user, developer). The study of these relationships under the viewpoint angle allows analyzing them globally, putting forward concerns related to human computer interface. Natural language is studied using statistic text processing techniques [21], [22] which aim at classifying information useful to visualize shared knowledge.

On the other hand, in order to maintain consistency among these viewpoints, RM-ODP puts a set of four rules categories:

- Basic rules,
- Object model rules. This category of rules provides the powerful concepts of multiple types that an object can assume, and multiple interfaces that an object can offer.
- Structuring and specification rules. These rules include organization, properties, naming, behavior, as well as abstraction, refinement, and composition concepts, which provide unique capabilities to architect a system.
- Conformance rules.

## 3. DBMS Object Model

The RM-ODP international standard [5] presents a very good architectural framework for modeling [26] distributed systems. However, nowadays, there are unfortunately not many modelers that use the standard in their everyday practice. It's a pity, considering the amount of highly qualified experts' knowledge invested in the project and the big constructive potential that its results might bring to practice if they were adequately used. one of the ways to promote the use of RM-ODP in formalization of its framework. The formalization requires a careful and attentive translation of the standard definitions into formal logical constructions, but once done it would allow creation of ODP-based software toolsets that could bring to modelers an "easy to be applied" version of the standard.

Generally, the term object model refers to the collection of concepts used to describe objects in an object-oriented specification (OMG CORBA), Object model [5] and RM-ODP object model [4]. It corresponds closely to the use of the term data-model in the relational data model. To avoid misunderstandings, the RM-ODP defines each of the concepts commonly encountered in object oriented models. It underlines a basic object model which is unified in the sense that it has successfully to serve each of the five ODP viewpoints. It defines the basic concepts concerned with existence and activity: the expression of what exists, where it is and what it does. The core concepts defined in the object model are object and action. An object is the unit of encapsulation: a model of an entity. It is characterized by its behavior [30] and, dually, by its states. Encapsulation means that changes in an object state can occur only as a result of internal actions or interactions.

An action is a concept for modeling something which happens. ODP actions may have duration and may overlap in time. All actions are associated with at least one object: internal actions are associated with a single object while interactions are actions associated with several objects.

Fig.1 shows depicting objects within a client system- server system community and depicting information objects data base management. It shows also many operators, each corresponding to one of the objects and each requiring services that relate to some part of the information schema. The information schema needs to have a shared and persistent representation, so a computational model of database systems interacting with the operators via their interfaces is depicted. These examples use a simple diagrammatic modeling notation which is not part of RM-ODP.

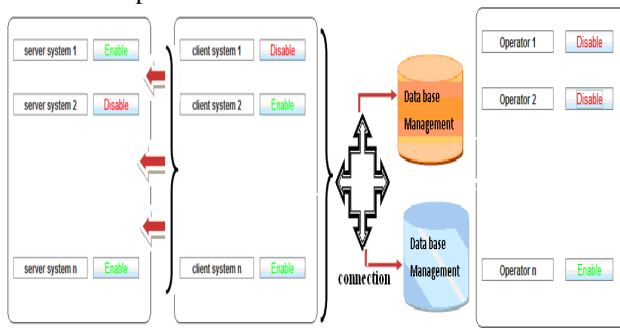

Fig.1: client system- server system interactions and data base management

## 4. Engineering Language and DMBS Structure

### 4.1. RM-ODP Common Functions

In addition to these structuring approaches, RM-ODP gives outline definitions of a number of common functions. Those functions [7, 8] provide a set of common services that are either fundamental or widely applicable to the construction of ODP systems. Detailed specifications for those functions are the subject of separate and specific standards.

RM-ODP also defines functions that are fundamental to the construction of any ODP systems. The functions are base architectural services that will be included in the implementation design.

The functions are organized into four groups' management, coordination, repository, and security:
1. Management functions :( node; object; cluster; capsule).
2. Coordination functions: (event notification; check pointing and recovery; deactivation and reactivation; group function; migration; engineering interface reference tracking and transaction function). These functions concern recording of event histories and ordering and notification of events and checkpointing objects, instantiating checkpoints, and undo or redo interactions for failure recovery. And for replication, they ensure coordination among replica objects and group membership management.
3. Repository functions: (storage; information organization; relocation; type repository and trading function), this is concerned with advertisement and discovery of interfaces.
4. Security functions: These functions ensure access control, authentication, security audit, key management, and confidentiality and integrity of information.

Those functions are forming an integral part of the computational language and an integral part of the engineering language. Most of them are introduced by the engineering language to provide the support needed for its structures.

Besides, the engineering language comprises concepts, rules and structures for the specification of an ODP system from the engineering viewpoint. Operating system and applications are an example of a node.

The Engineering held by the ODP system about entities in real world, including the ODP system itself, is modeled in an Engineering specification in terms of DBMS objects, and their relationships and behaviors.

### 4.2. DMBS Structure

Basic DBMS elements are modeled by atomic DBMS objects. More complex information is modeled as composite DBMS objects which, as any other ODP object, exhibit behavior, state, identity and encapsulation. Its type is a predicate characterizing a collection of DBMS objects, which their class is the set of all DBMS objects satisfying a given type.

Furthermore, an action is a model of something that happens in real world. Actions are instances; their types are modeled by ODP action types. An action in the information viewpoint is associated with at least one DBMS object Class. It can be either internal action or interaction as seen before.

DBMS objects template specifies the common features of a DBMS objects collection in sufficient detail that a DBMS objects can be instantiated using it. It may reference static, invariant and dynamic schema.

An **invariant schema** is a set of predicates on one or more DBMS objects which must always be true. The predicates constrain the possible states and state changes of the objects on which they apply. ODP also notes that an invariant schema can specify the types of one or more DBMS objects; that will always be satisfied by whatever behavior the objects might exhibit.

A **static schema** defines the state of one or more DBMS objects, at some point in time, subject to the constraints of any invariant schema.

A **dynamic schema** is a requirement of the allowable state changes of one or more DBMS objects, subject to the constraints of any invariant schema. A dynamic schema specifies how the information can evolve as the system operates. In addition to describing state changes, dynamic schema can also create and delete DBMS objects, and allow reclassifications of instances from one type to another. Besides, in the Engineering language, a state change involving a set of objects can be seen as an interaction between those objects. Not all the objects involved in the interaction need to change state; some of the objects may be involved in a read-only manner [29].

### 4.3. Syntax Domain

We define in this section the meta-models for the concepts presented in the previous section. **Fig.**2 defines the context free syntax for the Engineering language.

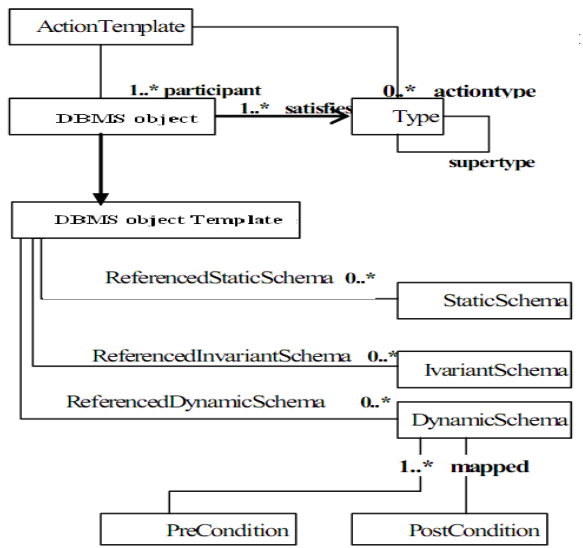

Fig.2 RM-ODP Foundation and DBMS concepts

In the following, we define context constraints for the defined syntax.
**Context m:** Model inv: m.Specifier->includes All (m.DBMS object Templates. Dynamic Schema)
m.Describer->includesAll (m. DBMS Template.StaticSchema)
m.Constrainer->includesAll (m. DBMS object.Invariant Schema)
m.ActionTemplates -> includesAll (m.DBMS object Templates.action)
m.Types->includesAll(m.ActionTemplates.

Types -> union (m.DBMS object.Types)
We consider the concepts of subtype/supertype (RM-ODP 2-9.9) and subclass/superclass (RM-ODP 2-9.10) as relations between types and classes respectively.
**Context m:** model inv: m.types-> forall( t1: Type, t2: Type | t2.subtype -> includes(t1) implies t1.valid_for.satisfies_type=t2)
m.types-> forall( t1: Type, t2: Type | t1.supertype ->includes(t2) implies t1.valid_for.satisfies_type=t2)
**Context a:** ActionTemplate inv: a.DBMS object.StartState <> a.DBMS object.EndState
**Context o:** Object Template inv: iot (DBMS object template) is not parent of or child of itself not (iot.parents ->includes (iot ) or iot.children->includes(iot)).

### 4.4. Semantics Domain

The semantics of a UML model is given by constraining the relationship between a model and possible instances of that model (see Fig.3). It means constraining the relationship between expressions of the UML abstract syntax for models and expressions of the UML abstract syntax for instances. We define a model to specify the ODP Engineering viewpoint: a set of DBMS objects, their relationships and behaviors. This model defines DBMS Semantic Domain (Fig.3).

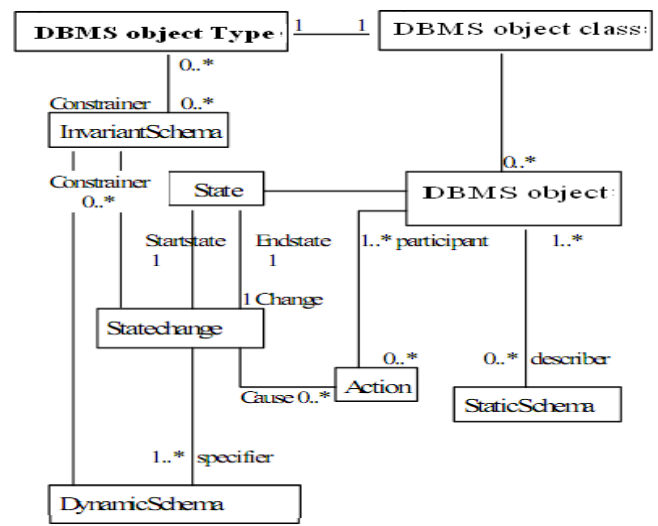

Fig.3 DBMS Semantic Domains

In addition, a system can only be an instance of a single system model, because it is self contained and disjoint from other models. On the other side, objects are instances of one ore more object templates; they may be of one or several types. With no further constraints, it is possible for an object to change the templates of which it is an instance; thus this meta-model supports dynamic types.

There is one well-formedness rule for instances, which

are given bellow:

**Context s:** system inv:

The source and target DBMS object of s'slinks is DBMS object in s:

s.DBMS objects->includesAll(s.links.source -> union(s.links.target))

Links between two DBMS objectare unique per role
s.links->forAll(l|s.links ->select (l'|l'.source=l.source&l'.target=l.target&l'.of=l.of)=l)

Declaration of "Specification concepts" (RM-ODP 2.9) in Alloy [28], time dependence.

**Context Time inv:** forall(o:DBMS object ,t:Time | t.instant ->notEmpty implies o.state ->notEmpty)

**Context Precondition inv:** Forall (prec: Dynamicschema.Precondition, o: DBMS object | exists(s: State) | o.mappedTo = prec and o.state_start = s)

**Context Postcondition inv:**forall (postc: dynamicschema.Postcondition , o :DBMS object | exists(s : State) | o.mappedTo = postc and a.state_end = s).

### 4.5. Meaning Function

Other invariants are required to constrain the relationships between models and instances. These constitute the semantics which are the subject of this section. The semantics for the UML-based language are defined by the relationship between a system model and its possible instances (systems). The constraints are relatively simple, but they demonstrate the general principle. Firstly, there are two constraints related to DBMS object and links, respectively.

The first shows how inheritance relationships can force a DBMS object to be of many DBMS object Template.

**Context o:** object inv:

The templates of o must be a single template and all the parents of that template

o.of->exists (t | o.of=t->union (t.parents))

The second ensures that a link connects objects of templates as dictated by its role.

**Context l:** link inv:

DBMS object which are the source/target of link have templates which are the source/target of the corresponding roles.

(l.of.source)->intersection (l.source.of) -> notEmpty and (l.of.target)->intersection(l.target.of)->notEmpty

Secondly, there are four constraints which ensure that a model instance is a valid instance of the model, it is claimed to be an instance of:

The first and second ensure that objects and links are associated with templates known in the model.

**Context s:** system inv:

The model, that s is an instance of, includes all object templates that s.objects are instances of.s.of. DBMS object Templates->includes All (s.DBMS objects.of)

The model, that s is an instance of, includes all DBMS object Class that s. DBMS Objects are instances of s.of.DBMS object Class ->includesAll(s.s. DBMS Objects.of).

The third ensures that links are associated with roles known in the model.

**Context s:** system inv:

The model, that s is an instance of, includes all the role that s.links are instances of

s.of.roles ->includesAll(s.roles.of)

The fourth constraint ensures that the system cardinality constraints on roles are observed.

**Context s:** system inv:The links of s respect cardinality constraints for their corresponding role

s.links.of -> forAll( r | let links_in_s be r.instances ->intersect ( s.links ) in ( r.upperBound -> notEmpty implies links_in_s ->size <= r.upperBound ) and links_in_s->size >= r.upperbound)

The fifth ensures that reverse links are in place for roles with inverses. If a link is of a role with an inverse, then there is a corresponding reverse link

s.links->forAll (l | l.of.role.inverse ->notEmpty implies s.links ->select (l' | l'.source=l.target & l'.target=l.source & l'.of = l..of.inverse) ->size=1.

## 5. DBMS Engineering Specifications.

An engineering specification defines the infrastructure required to support functional distribution of an ODP system by:
- Identifying the ODP functions necessary to manage physical distribution, communication, processing and storage.
- Identifying the roles of different DBMS object supporting the ODP functions.

In order to do this, we specify:
1. The activities that occur within those DBMS objects
2. The interaction of the DBMS objects (Fig.4).

To achieve that, we respect the engineering language rules like interface reference rules, binding rules, cluster, capsule and node rules, etc

### 5.1. DBMS Object Activities

The functions of a software entity are:
- Transferring a software entity;
- Creating a software entity;
- Providing globally unique operator names and locations;
- Supporting the concept of a domain;
- Ensuring a secure environment for software entity operations.

We specify these functions of a Software entity with the

ODP functions.

### 5.2. DBMS Object Interactions

We define three types of interactions related to interoperability:
- Remote software entity creation;
- Interaction needed for the software entity transfer;
- Software entity method invocation.

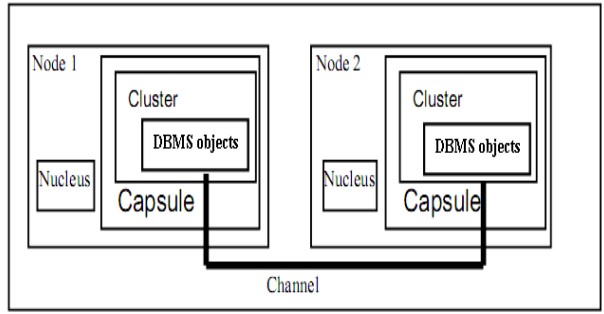

Fig.4: organization of the DBMS objects

A client could be a non-software entity program or a software entity from a software entity having the same system type as the destination agent or not. This client authenticates itself to the destination software entity system and interacts with the destination software entity to request the creation of a software entity.

When a software entity transfers to another software entity, the software entity system creates a travel request providing information that identifies the destination place. In order to fulfill the travel request, the destination software entity transfers the software entity's state, authority, security credential and the code.

For example in database System server, a channel between system client manager Object and the system server Object can be defined as illustrated in Fig.5.

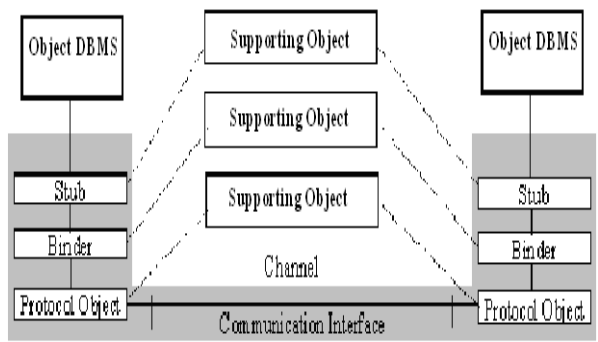

Fig.5: An example of a basic system client / system server channel.

A system client object invokes a method of another system client object or system server object if it has the authorization and a reference to the system client object.

### 6. Conclusion.

In the present paper, we have seen that the Reference Model for Open Distributed Processing (RM-ODP) provides a framework which supports the integration of distribution, interworking and portability. In addition, the UML standard has adopted a meta-modeling approach to define the abstract syntax of UML. One approach to define the formal semantics of a language is denotational: essentially elaborating the value or instance denoted by an expression of the language in a particular context. However, the ODP viewpoint languages define what concepts should be supported and not how these concepts should be represented, So when we used the denotational meta-modeling approach in this paper, we have defined the UML/OCL based syntax and semantics of a language for a fragment of ODP object concepts described in the foundations part and in the Engineering viewpoint language. Indeed, these concepts are suitable to define and constrain ODP Engineering viewpoint specifications. In parallel, we have applied the same approach to define a language of DBMS concepts characterizing dynamic behavior of DBMS objects.

### 7. References

[1] ISO/IEC, Basic Reference Model of Open Distributed Processing-Part1: Overview and Guide to Use, ISO/IEC CD 10746-1, July 1994.
[2] ISO/IEC, RM-ODP-Part2: Descriptive Model, ISO/IEC DIS 10746-2, February 1994.
[3] ISO/IEC, RM-ODP-Part3: Prescriptive Model, ISO/IEC DIS 10746-3, February 1994.
[4] ISO/IEC, RM-ODP-Part4: Architectural Semantics, ISO/IEC DIS 10746-4, July 1994.
[5] OMG, the Object Management Architecture, OMG, 1991. http://www.omg.org
[6] Jalal Laassiri and al."A Denotational Semantics of Concepts in ODP Information Language", Proceedings of the International MultiConference of Engineers and Computer Scientists , London, UK, July 2009,pp 486-492
[7] ISO/IEC, ODP Type Repository Function, ISO/IEC JTC1/SC7 N2057, January 1999.
[8] ISO/IEC, the ODP Trading Function, ISO/IEC JTC1/SC21, June 1995.
[9] J.M. Spivey, The Z Reference manual, Prentice Hall, 1992.
[10] IUT, SDL: Specification and Description Language, IUT-T-Rec. Z.100, 1992.
[11] ISO and IUT-T, LOTOS: A Formal Description Technique Based on the Temporal Ordering of Observational Behavior, ISO/IEC 8807, August 1998.
[12] H. Bowman et al. FDTs for ODP, Computer Standards & Interfaces Journal, Elsevier Science Publishers, Vol.17, No.5-6, 1995, pp. 457-479.
[13] J. Rumbaugh et al., The Unified Modeling Language, Addison Wesley, 1999.

[14] B. Rumpe, A Note on Semantics with an Emphasis on UML, Second ECOOP Workshop on Precise Behavioral Semantics, Technische Universitaty unchen publisher, 1998.

[15] A. Evans et al., Making UML precise, OOPSLA'98, October 1998,

[16] Evans et al. The UML as a formal notation, UML'98, France June 1998, LNCS 1618, Springer Berlin, 1999, pp. 336-348

[17] J. Warner and A. Kleppe, the Object Constraint Language: Precise Modeling with UML, Addison Wesley, 1998.

[18] M. Bouhdadi et al, An UML-based Meta-language for the QoS-aware Enterprise Specification of Open Distributed Systems, IFIP TC5/WG5.5 Third Working Conference on Infrastructures for Virtual Enterprises (PRO-VE'02), May 1-3 Sesimbra Portugal, Kluwer Vol. 213 (IFIP Conference Proceeding series), 2002.

[19] Jacques Saraydaryan and al, Comprehensive Security Framework for Global Threats Analysis, IJCSI Volume 2, August 2009.

[20] S. Kent, S. Gaito, N. Ross. A meta-model semantics for structural constraints in UML,, In H. Kilov, B. Rumpe, and I. Simmonds, editors, Behavioral specifications for businesses and systems, chapter 9, pages 123-141. Kluwer Academic Publishers, Norwell, MA, September 1999.

[21] D.A. Schmidt, Denotational semantics: A Methodology for Language Development, Allyn and Bacon, Massachusetts, 1986.

[22] Myers, G. The art of Software Testing, John Wiley &Sons, New York, 1979

[23] Binder, R. Testing Object Oriented Systems. Models. Patterns, and Tools, Addison-Wesley, 1999

[24] Cockburn, A. Agile Software Development. Addison-Wesley, 2002.

[25] Bernhard Rumpe. Agile Modeling with UML. Habilitation Thesis, Germany, 2003.

[26] Beck K. Column on Test-First Approach. IEEE Software, 18(5):87-89, 2001

[27] Briand L. and Labiche Y. A UML-based Approach to System testing. In M. Gogolla and C. Kobryn (eds): "UML" – The Unified Modeling Language, 4th Intl. Conference, LNCS 2185. Springer, 2001 pp. 194-208,

[28] Bernhard Rumpe, Executable Modeling with UML. A vision or a Nightmare? In Issues & Trends of Information Technology Management in Contemporary Associations, Seattle. Idea group Publishing, Hershey, London, pp. 697-7001. 2002 Author, Title of the Book, Publishing House, 200X.

[29] A.Naumenko, A.Wegmann, "Proposal for a formal foundation of RM-ODP concepts » conference woodpecker 2001.

[30] ISO/IEC, May 2006, Basic Reference Model of Open Distributed Processing-Use of UML for ODP system specifications, ISO/IEC CD 19793.

[31] Naumenko, A., et al. A Viewpoint on Formal Foundation of RM-ODP Conceptual Framework, Technical report No. DSC/2001/040, July 2001, EPFL-DSC ICA.

[32] Wegmann, A. and A. Naumenko. Conceptual Modeling of Complex Systems Using an RM-ODP Based Ontology. in 5th IEEE International Enterprise Distributed Object Computing Conference - EDOC 2001. 2001.

**Jalal Laassiri:** Received the B.S.and M.S. degree from the University of Mohamed V, Morocco, in 2002 and 2005 respectively. He is a Ph.D. student at the department of Computer Science of the Faculty of Sciences Rabat Morocco. His research interests include formal verification techniques and performance evaluation methods of concurrent and distributed systems with applications to computer and communication systems, In 2009, he received the Best Student Paper Awards of the WCE 2009.Is a member of the IAENG, Is a member of the ICIIC2010 International Program Committee.